# Probing physics students' conceptual knowledge structures through term association

Ian D. Beatty[a)] and William J. Gerace

Physics Department and Scientific Reasoning Research Institute,
University of Massachusetts, Amherst, Massachusetts 01003-4525

**ABSTRACT**

Traditional tests are not effective tools for diagnosing the content and structure of students' knowledge of physics. As a possible alternative, a set of term-association tasks (the "ConMap" tasks) was developed to probe the interconnections within students' store of conceptual knowledge. The tasks have students respond spontaneously to a term or problem or topic area with a sequence of associated terms; the response terms and time-of-entry data are captured. The tasks were tried on introductory physics students, and preliminary investigations show that the tasks are capable of eliciting information about the stucture of their knowledge. Specifically, data gathered through the tasks is similar to that produced by a hand-drawn concept map task, has measures that correlate with in-class exam performance, and is sensitive to learning produced by topic coverage in class. Although the results are preliminary and only suggestive, the tasks warrant further study as student-knowledge assessment instruments and sources of experimental data for cognitive modeling efforts.

## I. INTRODUCTION

### A. Motivation

There is a growing consensus among educational researchers that traditional problem-based assessments are not reliable tools for diagnosing students' knowledge and for guiding pedagogical intervention, and that new tools grounded in the results of cognitive science research are needed. If one wishes to assess a student's state of knowledge, rather than merely summarize the parts of the assessment in which the student did and did not succeed, one needs a model of what a knowledge state is and how it is probed by the assessment. An effective diagnostic assessment must describe a student with reference to some suitably detailed model of physics knowing, learning, and application.

No sufficiently specific model of knowledge structuring and accessing yet exists to serve as a basis for detailed diagnostic assessment of conceptual understanding; it has been said that "knowledge representation is one of the thorniest issues in cognitive science."[1] Nevertheless, physics education research (PER) has provided general qualitative descriptions of knowledge structuring in physics that can help direct the search for new assessment approaches. Cognitive scientists distinguish between two fundamental kinds of knowledge: *declarative* and *procedural* knowledge.[1] In essence, declarative knowledge is explicit knowledge of facts, which can be stated or reported, and *procedural* knowledge is tacit knowledge of how to perform operations, which can be demonstrated but not stated. PER studies on physics experts' and novices' problem-solving behavior suggest that at least within the domain of physics, declarative knowledge can be divided into four general, approximate categories:[2-6] *conceptual* knowledge, *operational and procedural* knowledge, *problem-state* knowledge, and *strategic* knowledge. (The operational and procedural category refers to declarative knowledge about physics operations and procedures, as distinct from automated, non-declarative "procedural knowledge." The choice of terminology is unfortunate, especially because many operational skills have both declarative and procedural components.)

Further studies demonstrate that experts and novices are distinguished not just by the content of their knowledge stores, but by their organization:[7] experts have contextually-appropriate *access* to and not just possession of knowledge;[8] it is the *structure of interconnections* between knowledge elements that allows such access,[9] and experts' knowledge is structured around key *principles.*[10] These findings suggest that for purposes of assessing students' degree of expertise with respect to a physics topic, a need exists for tools that can probe students' declarative knowledge state in terms of the knowledge elements present and especially the *structure of interconnections* between those elements: the students' *conceptual knowledge structure*.

**B. Previous Approaches to Assessing Conceptual Knowledge Structure**

One device that has been developed for probing a student's conceptual knowledge structure is the *concept map.*[11,12] In a typical concept map assessment, a student is asked to draw a nodes-and-links representation of her understanding of a domain topic area. The resulting map is taken as a description, perhaps partial, of the student's declarative knowledge structure for the topic. Many variants have been proposed: sometimes the subject is asked to draw the entire map without assistance, with labeled or unlabeled links; sometimes he is given a set of terms to arrange into a map; sometimes a partial map is provided, and he is asked to fill in the remainder; and sometimes a complete map without link labels is given and the student is asked to label all links. Scoring systems also vary widely, with credit given for the number of nodes, the number of links, the number of nodes or links deemed relevant, the degree of similarity to a reference map, or some combination of these possibilities.

Concept maps have proven useful for educational research. Empirical evidence is mixed on the extent to which concept map-based measures of student knowledge correlate with other indicators such as standardized exam performance, perhaps due to the plethora of task formats and scoring systems investigated.[12-14] It is clear, however, that such assessments tend to be tedious and time-consuming to administer and to score and analyze, rendering them poorly suited for mass adoption by educators.[12,15] Some researchers have implemented concept map assessments by computer and automated the scoring procedures,[16] but so far no widely adopted assessment tools have resulted, perhaps because of doubts about the scoring protocols chosen.

Whether or not students are capable of drawing a concept map that accurately describes their actual knowledge structure is open to significant doubt. One reason for doubt is the observation that drawing a concept map is a time-consuming and attention-intensive activity, and a student is unlikely to be able to draw a map of any completeness for more than a very small set of concepts. In an attempt to probe students' domain knowledge more thoroughly and to capture information about the relative strengths of inter-concept links as well as the presence or absence of such links, inferred approaches to declarative knowledge assessment have been developed. One is the *item relatedness judgment* task,[17,18] in which students are presented with all possible pairings from a list of terms, one pairing at a time, and asked to rate the "relatedness" of each pair on a numerical scale. The result is a *proximity matrix* capturing information about the student's knowledge structure; the matrix is then analyzed in an attempt to reveal that structure, perhaps via a scaling procedure like *cluster analysis* or *multidimensional scaling,*[19,20] or perhaps via a network-construction algorithm like *Pathfinder.*[17-19,21]

Overall, investigations into the validity of such inferred approaches to declarative knowledge structure assessment have been generally positive: measures comparing the similarity of students' derived structures (networks or scaling procedure results) to experts' referent structures correlate significantly, though not completely, with more traditional measures of domain mastery.[17,18,21] Unfortunately, item relatedness judgment tasks must either confine themselves to small sets of terms or take an impracticably long time to administer, because the time required scales as the square of the number of terms included.

In addition, it is not obvious that a student's reflected judgment on two terms' relatedness necessarily corresponds to her implicit knowledge structure as it affects knowledge application. One possible reason is that such assessments impose a set of terms on the student, rather than drawing out the set that the student has unprompted access to (as some versions of the concept map approach can) — similar to testing someone's *passive vocabulary* rather than her *active vocabulary* in a language. Another is that the student may be able to appreciate that two given terms are related, if asked, but not have the relation come to mind when needed for use in a problem-solving context.

**C. ConMap**

As a step toward the development of practical tools for assessing physics students' conceptual declarative knowledge structuring, we have developed a set of brief computer-administered tasks for eliciting students' conceptual associations.[22] The tasks are collectively referred to as *ConMap* ("conceptual mapping") tasks. The basic approach of the tasks is to elicit spontaneous term associations from subjects by presenting them with a prompt term, or problem, or topic area, and having them type a set of response terms. Each response is recorded along with the time spent thinking of and typing it, in an attempt to capture the flow of concepts triggered in each subject's mind. The specific tasks and their administration will be described in more detail below. (The traditional hand-drawn concept map task, which we use for comparison, is *not* a ConMap task.)

To investigate the information that the ConMap tasks might reveal about students' knowledge structuring, several studies were conducted between 1997 and 1999 with subjects from various introductory physics courses taught at the University of Massachusetts. Many different aspects of the data were analyzed, including extensive statistical treatment of the timing data associated with each term response list. This paper will consider results primarily from one particular study, and present selected analysis of the term lists without reference to the accompanying timing data. For detailed descriptions of all studies and thorough presentation and discussion of all analysis, see Refs. 22 and 24.

The purpose of this paper is *not* to display the ConMap tasks as finished practical assessment tools, but rather:

(1) To introduce the tasks as suggestions for a style of assessment that might eventually be useful and complement existing assessment approaches;

(2) To present evidence that the various tasks are, at least to some degree, sensitive to the aspects of knowledge and learning that we wish to probe; and

(3) To share some intriguing phenomena exhibited by the task results.

## II. THE STUDY

### A. The ConMap Tasks

Several brief, computer-administered tasks were developed to elicit spontaneous conceptual associations. To probe the conceptual portion of declarative knowledge, most of the ConMap tasks attempt to elicit subject's associations between *terms*. The focus is on terms rather than on equations, propositions, or other kinds of entities because terms seem to be the closest accessible approximation to "conceptual building blocks." This paper is not concerned with the underlying cognitive nature of such building blocks, or with the neurological details of their representation, storage, and retrieval.

It has proven difficult to rigorously define *term*. When instructing subjects, a term was loosely defined to be one or perhaps two or three words describing one concept, idea, or thing. Some examples of terms drawn from introductory mechanics are "kinematics,"

“Newton’s first law,” “pulley” and “problem-solving.” Statements like “energy is conserved in an elastic collision” were not considered to be terms, but rather propositions involving multiple terms and their relationship. “Conservation of energy,” on the other hand, would be accepted as a term, because it serves as a name for a physics concept. In practice, the distinction between single-concept terms and compound statements of relationship is not sharp, and subjects frequently wandered dismayingly far over it.

Knowledge is context-dependent, in the sense that the knowledge accessible to a student depends on the student’s current cognitive context. Has the student been asked a question about work? Is she thinking about a problem involving an inclined plane? Is he reviewing his physics course to date, perhaps chronologically? Therefore, several different ConMap tasks have been developed, each intended to specify a context for the subject in a different manner and therefore probe a somewhat different aspect of the subject’s knowledge store.

One task was the *Term Prompted Term Entry* (TPTE) task, in which subjects were given a prompt term from the physics domain. They were asked to think of terms they consider related to this prompt term, rapidly, spontaneously, and without strategy, and to type these terms into a dialog box (Fig. 1) as the terms came to mind. The prompt term stayed visible throughout, and typed terms disappeared from view as they were entered. Data gathered for each subject consists of the response terms, together with the time at which typing began for each response (the moment at which the first character was typed into an empty field), and the time at which each response was completed (the moment at which the return key was pressed).

For a given prompt term, the task was terminated after ten terms were entered, or the first time the subject paused with an empty response box for more than ten seconds (if at least three terms had been entered at that time). The process was repeated for several different prompt terms. This task was intended to probe, as directly and free of context as possible, the immediate conceptual neighborhood of specific concepts.

A second task investigated was the *Problem Prompted Term Entry* (PPTE) task. This task was identical to the TPTE task in all respects, except that the prompt was a physics problem or problem situation rather than a term. Subjects were instructed by the computer when to turn the page in a ring binder, revealing the new prompt problem. They then read the problem on paper and began entering responses into a dialog box (like the TPTE dialog in Fig. 1, but without the prompt term). This task was intended to explore the conceptual associations inherent to the subject in the context of a specific physics problem.

A third task investigated was the *Free Term Entry* (FTE) task. For this, subjects were prompted with a general topic area like “introductory mechanics” or “the material covered in your physics course this semester,” and asked to enter terms spontaneously, as they came to mind, for the duration of the task (typically 20 to 45 minutes). Subjects were specifically directed to enter as many terms as possible from within the specified topic area, and to persevere to the end of the task. The task differed from true free association in that subjects were instructed to refrain from entering terms outside the designated topic area, and to avoid entering any given term more than once if possible. This task was intended to broadly survey a subject’s structuring of a conceptual domain or topic area, with no detailed context (such as a problem to be solved) to shape or filter the subject's perception of the topic.

In addition, a traditional paper-and-pencil *Hand Drawn Concept Map* (HDCM) task was included in the study, for comparison with the other tasks. Subjects were given a prompt term like "energy" and instructed to draw a concept map around that term. Subjects were free to select their own terms, links between nodes were not to be labeled, and the map structure need not be hierarchical. Subjects were to continue elaborating their map for the duration of the task (typically 10 or 12 minutes).

Ultimately, a combination of these and perhaps other tasks might make it possible to construct a reasonable representation of a physics student's conceptual knowledge structure, including information about how access to that knowledge store is limited and constrained by context. This paper, however, is merely concerned with establishing whether the tasks are capable of probing knowledge structure at all.

**B. Study Design**

During the spring semester of 1999, volunteers were solicited from the Physics 151 course (introductory mechanics for science and engineering majors) at the University of Massachusetts at Amherst, shortly after the first of four course exams. Financial compensation was offered. Sixteen subjects were chosen from the volunteer pool, representing both genders and a range of exam 1 scores from C to A. All subjects were native English speakers.

Each subject participated in nine 15-minute sessions and one final 90-minute session, scheduled weekly for the remainder of the semester. Sessions were run under controlled conditions and monitored. One or two tasks were conducted during each 15-minute session, and four tasks, a group interview, and profile questionnaire were given during the 90-minute final session. The TPTE task was given during eight of the sessions, with a total of 52 prompts (28 unique, most repeated twice during the study). The PPTE task was given during seven of the ten sessions, with a total of 40 prompt problems (30 unique); five of the prompt problems were "problem situations" with no associated question. The HDCM task was given during four sessions, with three unique prompt terms.

## III. ANALYSIS

For each prompt term of each TPTE and PPTE task, the data obtained consisted of a list of response terms (no more than ten) and associated timing information. Analysis of the timing data will not be discussed in this paper. For each HDCM task, the drawn map was the only source of data. Every term appearing on a map was classified according to its *level*, indicating how far removed it was topologically from the map's prompt term. For example, a term directly linked to the prompt term was classified as level 1, while a term directly connected to a level 1 term but not to the prompt term was classified as level 2.

The following subsections present results from some specific analyses performed on the study data.

## A. TPTE versus HDCM

This section compares subjects' hand-drawn concept maps (HDCM) to their term-prompted term entry (TPTE) response term lists for identical prompts. During session B of the study, a TPTE task was given in which "force" was one of the prompt terms presented, and a HDCM task was given with "force" as the prompt term. Similarly, session H included TPTE and HDCM tasks with the prompt "momentum," and session J included TPTE and HDCM tasks with the prompt "force." Session G included a HDCM task with the prompt "energy;" although that session did not include a TPTE task, energy was used as a TPTE prompt during session J. Thus, there were three occasions on which a prompt term was used for both a TPTE and HDCM task during the same session, and one on which the TPTE and HDCM tasks were separated by three weeks.

In all sessions that included a HDCM task, that task was placed at the end of the session, after all term-entry style tasks. This was done because the term-entry tasks are by design rapid and spontaneous, relying on impulsive associations, while the drawing of a concept map by hand is a much slower, more contemplative and reflective task. It therefore seemed likely that the term-entry tasks would be more susceptible to "pollution" from prior tasks: even if previously-used term-entry responses occurred to students during construction of a concept map, they had the time and freedom to ponder whether those terms belonged in the map. No empirical data was obtained on how the relative ordering and temporal separation of the TPTE and HDCM tasks impacts the responses. Extensive investigation of this impact is clearly important before any practical assessments are attempted.

For each of these four pairings, each of the sixteen subjects' HDCM maps was compared to his or her TPTE response list, resulting in 64 map/list comparisons. For each map/list comparison, each term in the TPTE response list was matched with an HDCM node containing an equivalent term, if one existed, and the level of that node was noted. If the TPTE response list contained duplicate terms, repeats were ignored. Because subjects were free to choose their own phrasing and spelling, inexact matches were common, so a TPTE response term was defined to match a map node term if their meanings were equivalent whether or not the terms were identical. For example, "gravity" and "gravitation" were considered matches, as were "$F_N$" and "normal." Contextual clues from adjacent nodes were sometimes used to aid in identifying the intended meaning of map terms. On occasion, a TPTE response term did not appear by itself as a map node, but did appear as part of a compound term in a map node: for example, "acceleration" might appear in the TPTE response list and not on the HDCM map, but "mass □ acceleration" might appear on the map. In such cases, the term was counted as appearing on the map, with the level of the compound term containing it.

For each map/list pair, the fraction of level 1 map terms that appeared in the corresponding TPTE response list was calculated. The mean and standard deviation of this fraction across study subjects are displayed in Table I. On average, slightly more than half of the terms from each subject's first-level map nodes also appear in the subject's corresponding TPTE response list. This overlap value was atypically low for a few map/list pairings; when such pairings were inspected in detail, it was often found that the subject had "categorized" several of the terms from the TPTE response list and used that category as a first-level node on the HDCM, causing the terms themselves to appear at the second level. For example, a subject might have listed several kinds of forces as TPTE responses to the prompt “force,” but might have categorized kinds of forces into “contact” and “at a distance” on the HDCM, with the two category names directly linked to the central “force” node and the specific forces connected at level two.

For each map/list pair, the fraction of TPTE response terms not appearing anywhere on the map was calculated. Between 0% and 35% of a subject’s TPTE response terms are typically absent from his corresponding HDCM. Table II displays the mean and standard deviation of this fraction across study subjects.

Despite the fact that the HDCM is a considered, reflective task and the TPTE is a spontaneous, impulsive one, TPTE data sets seem to provide a subset of the information provided by a HDCM. Specifically, a subject’s TPTE response list typically contains slightly more than half of the first-level terms appearing in the corresponding HDCM, and few of the TPTE responses are entirely absent from the HDCM. The TPTE thus seems useful for probing the core structure of a subject’s conceptual knowledge store, while the HDCM gathers more widespread structural information.

**B. TPTE Response Scoring**

In a preliminary attempt to investigate whether the TPTE task could serve as a useful student assessment tool, a procedure was developed for assigning a score to a subject’s TPTE response list based on the quality of the response terms entered as judged by domain experts. The resulting set of scores was compared against subjects’ performance on their in-class exams. Scoring of lists was carried out for only one prompt term, "force," which was used as a TPTE prompt during sessions B, C, and J of the study.

A panel of five physics experts — four physics professors and one advanced graduate student — was formed. Four of the five had detailed knowledge of the ongoing ConMap research project, so the panel cannot be considered representative of any general population of physics experts. To familiarize the expert panelists with the TPTE task and to acquire some data for later comparisons, the experts were all assigned a 16-prompt TPTE session which included the prompt "force."

A master list was constructed which consisted of every response term given by every study subject to the TPTE prompt "force" in each of the three sessions in which it was presented, and also every response term given by each of the five expert panelists to the prompt "force." Terms that were only trivially different representations of the same concept (for example, "conservation of energy" and "energy conservation") were mapped to one standard version, resulting in a set of 80 terms. This set was alphabetized and presented to each of the expert panelists. The experts were instructed to rate the quality of each term as a TPTE response to the prompt "force," and assign to it a 2, 1, or 0, according to the following scale:

2: Good/valuable/important. "This student knows his/her stuff."

1: Has some merit. "Not an unreasonable response."

0: Irrelevant, worthless. "Reveals no nontrivial knowledge."

The five experts' ratings were averaged for each response term, resulting in a quality value between 0 and 2 for that term relative to the prompt term.

The score for a subject's response list was defined to be the sum of the quality values for each term in that list; a list score can therefore range from 0 to 20. Such a score was calculated for each of the subjects' response lists to the prompt "force" in each of the three sessions. The resulting set of 48 scores (three each for sixteen subjects) ranged from 4.2 to 15.2, with a mean of 11.8 and a standard deviation of 2.9. For each subject, a mean score for the "force" prompt was calculated by averaging her scores for each of the three sessions' response lists. The resulting set of 16 mean scores ranged from 8.0 to 14.5, with a mean of 11.8 and a standard deviation of 2.0.

A scatterplot of each subject's mean response list score to "force" versus the sum of their raw scores on the four course exams — a crude measure of overall subject expertise — is shown in Fig. 2. The Pearson's *r*-value coefficient of correlation was 0.608, where 0.13 is the threshold for statistical significance with 16 data points. (The coefficient of correlation is defined to be statistically significant if it implies that the relative error in the slope of the best-fit line is less than 1/3.[23]) This result suggests that the TPTE response list scores calculated according to the above procedure correlate with subject expertise as measured by exam scores. A different scoring rubric might of course produce a stronger correlation, as might comparing TPTE scores to a more targeted measure of exam performance based only on force-related questions. These hypotheses have not been tested, but the simple rubric and comparison performed suffices to demonstrate the existence of a correlation.

## C. PPTE Response to Course Coverage

Two Problem-Prompted Term Entry (PPTE) prompt problems were given during three different sessions of the study, and four others were given during two different sessions. In an attempt to determine whether subjects' PPTE responses were impacted by their learning of the subject material during the associated physics course, their responses from different sessions for the same prompt problem were compared. The notation "problem D6" means the sixth prompt problem given during the session D PPTE task. All subjects were given prompt problems in the same order, so problem D6 was identical for all subjects.

Two of the problems given two times each (I2 = J4 and I5 = J1) were separated by only one week, and both sessions were significantly later in the semester than the relevant material was covered. These two cases served as a control test by providing a measure of how consistent subjects' PPTE responses were for two consecutive sessions, in the absence of directly relevant course coverage.

The two problems given three times each (C1 = D6 = J3 and C4 = D2 = J5) were given during sessions C, D, and J. Domain material relevant to the problems was covered in the concurrent physics course between sessions C and D, and an exam on the material was given during the same week as session D, so it is reasonable to assume that subjects spent time studying the material during the week between sessions C and D. The data from these two prompt problems were examined as a test of the hypothesis that the PPTE task can detect conceptual change resulting from lecture coverage and exam studying.

Comparison with the session J responses served as an additional control, to test whether any apparent change between C and D was long-lasting or temporary.

1. <u>Control: Consecutive Weeks, No Course Coverage</u>

PPTE prompts I2 and J4 used the same prompt problem (taken from the third course exam, given during the week before session H). The same is true of prompts I5 and J1. For each occurrence of each prompt, subjects who responded with terms indicating the key concept(s) needed to solve the problem were identified, and a comparison was done to see how consistent subjects were in this regard across the two sessions.

The problem used for I2 and J4 is shown in Fig. 3. For both presentations, each subject who included "momentum" or "conservation of momentum" among his responses was binned as "positive," regardless of what other responses were included. The purpose of this scheme was to detect whether the relevant concept was brought to the subject's consciousness by the problem, not to determine whether the subject could select it from among other concepts. Thirteen of the subjects were positive (included the concept) for both I2 and J4; two were negative for both; and only one changed categories, from negative to positive.

Figure 4 shows the problem used for I5 and J1. For both, a subject was binned as positive if he included "momentum," and also included "conservation of energy," "conservation of mechanical energy," or both "kinetic" and "potential" energy. As with the comparison of I2 and J4, only one subject in sixteen changed categories, from negative in I5 to positive in J1 (not the same as the lone category-changing subject in the previous comparison); twelve were positive for both and three were negative for both. The one subject who changed categories was a marginal negative for I5, and an argument could be made for placing him or her in the positive bin, which would mean no subjects changed bins at all.

The results of these two comparisons suggest that in the absence of direct course coverage of the relevant subject material, the likelihood that students will respond with a PPTE term relevant to the prompt problem's solution is approximately the same when the task is given in two consecutive weeks: no coverage, no change.

2. Lecture Coverage and Exam Studying

The two problems given during sessions C, D, and J are optimally solved with the work-energy theorem. When subjects were presented with the problems during session C, they had been introduced to work and energy concepts, but the lecture instructor had not completed his treatment of the work-energy theorem. During session D, one week later, coverage of energy topics was essentially complete, and subjects were taking an exam on the material. The session J presentation occurred significantly later, at the end of the semester.

It was hypothesized that additional lecture and homework coverage of the material and preparation for the exam would impact the way subjects responded to the prompt problems. Specifically, it was anticipated that more students would respond with terms indicating an inclination to consider the work-energy theorem for solving the problems during session D than during session C. For the session J responses, two outcomes seemed plausible, assuming that the hypothesis about sessions C and D turned out to be correct: if the increase from C to D was due to short-term immersion in work and energy course material (that is, subjects had those terms on their minds), then the fraction of positively-binned subjects should decrease from D to J; or, if the increase was due to a real change in subject's conceptual reaction to the problems, then the rate for J should be comparable to the rate for D and significantly higher than the rate for C.

Problem C1 had no picture, and read: "An object is launched directly upward with an initial speed of 18 m/s. What is the object's speed after rising 8 meters?" Problem D6 was identical to C1 except that "18 m/s was" changed to "12 m/s" and "8 meters" was changed to "5 meters". Problem J3 was identical to problem D6.

Each subject was binned as positive if he included "work" or "energy" as a response term or part of a response term in his list of responses to the prompt problem. Subjects who mentioned neither were binned as negative. Each subject was binned according to this criterion for sessions C, D, and J, resulting in three binnings per subject. 1 of 16 subjects were binned as positive for C1, 7 of 16 for D6, and 6 of 16 for J3. When comparing subjects' binnings for D6 and J3, it was found that seven were negative for both sessions, four were positive for both, two changed from negative in D6 to positive in J3, and three changed from positive in D6 to negative in J3.

Problem C4 had no picture, and read: "A 30 kg box starts from rest on a frictionless horizontal floor. A force of 200 N is applied to the box, pushing down at an angle of 45˚. How much work must the applied force do to get the box moving at 1 m/s?" Problem D2 was identical except that "30 kg" was changed to "25 kg", "200 N" was changed to “320 N”, and “1 m/s” was changed to “1.5 m/s”. Problem J5 was identical to problem D2.

For each subject’s session C list of response terms to the prompt problem, the subject was binned as positive if she included “energy” or “work-energy theorem” as a response term or part of a response term. Subjects who mentioned neither term were binned as negative. Subjects who merely entered “work” were binned as negative, because the problem itself explicitly asks for the work to be determined. 3 of 16 subjects were positive in C4, 6 of 16 for D2, and 6 of 16 for J5. When comparing subjects’ binnings for D2 and J5, it was found that eight were negative for both sessions, four were positive for both, two changed from negative in D2 to positive in J5, and two changed from positive in D2 to negative in J5.

These two comparisons suggest that when presented with a PPTE problem, students are more likely to include among their responses a term indicative of the concept necessary for the problem’s solution after they have been exposed to material containing the concept through lecture, homework, and studying. When such exposure occurred, positive responses increased noticeably over one week, and remained higher six weeks later. Also, subjects’ binnings remained relatively stable for the six weeks following the exposure: two-thirds to three-quarters of the subjects were in the same bin for both the session D and session J prompts, for both comparisons. Overall, there appears to be suggestive evidence that the PPTE task is sensitive to course-induced learning. How sensitive it is, and how influenced it might be by details of the course, has not been determined.

One might ask why only seven of sixteen students, all of whom scored a C or better on the first course exam, mentioned “work” or “energy” in response to problem D6 after a week of intensive lecture, homework, and studying on work and energy topics. C1/D6/J3 is, after all, a rather straightforward conservation of energy problem. If the PPTE task is providing information about the subjects’ ability to solve the problem, this response would suggest that the subjects are largely unable to apply their recently-acquired knowledge to even simple problems. Subjects were not asked to actually solve the problems during the study, so no direct data exists to resolve this question. However, it is suggestive to note that twelve of the sixteen subjects included “kinematics” or a clear equivalent among their responses to D6, including seven of the nine who included “work” or “energy,” and five of the seven who didn’t. Many of the subjects had “kinematics” near the beginning of their response list. It would seem that despite their recent immersion in work and energy, the subjects retained a strong inclination to react to the problem as an exercise in kinematics, perhaps because of longer experience and greater confidence with that topic.

It is also worth remembering that subjects were not instructed to enter terms indicative of the prompt problem’s solution, but merely terms they considered related to the problem, spontaneously and without reflection. The task’s instructions could be rewritten so that subjects enter terms related to the solution of the problem, but this would destroy the spontaneity of the associations elicited. And even then the task shouldn’t be expected to accurately predict problem-solving success: a student might very likely (for example) begin problem D6 with kinematics, get frustrated, and then turn to energy concepts and succeed; the PPTE task would elicit kinematics only, unless the subject thought through their solution completely and clearly before entering terms. Recall that the task is not intended as a replacement for problem-solving assessments as a measure of problem-solving ability, but rather as a probe of the linkages between students' *conceptual* and *problem-state* knowledge.

**D. FTE Jump Rates**

When a subject generates terms for a Free Term Entry (FTE) task, it seems plausible that he "walks" the network of concepts in his knowledge structure representing the given domain, stepping from concept to concept along relatively strong links that associate concepts. If this is true, adjacent terms in FTE response lists should be generally related, with occasional jumps when no associated terms immediately suggest themselves to the subject and he has to stop and think for a while to come up with another not-yet-entered term. (This is of course a simple model, ignoring other possible mechanisms such as parallel processing and delayed subconscious processing.)

One testable hypothesis that follows this picture is that when two adjacent terms in a FTE response list are strongly related, the elapsed time between entering the two should be smaller on average than the time between two relatively unrelated terms. This will be investigated in a subsequent paper.[24] If it is assumed that more expert-like (that is, better-performing students) have more richly structured conceptual knowledge, then another testable hypothesis is that subjects' course grades should correlate with the fraction of jumps in their FTE response lists. This subsection tests that hypothesis.

Define a term to be a jump if it does not appear to be reasonably related to one of the preceding three terms in the subject's sequence of FTE responses, according to a domain expert's judgment. The three-term threshold was chosen because according to the introspective testimony of experts who were given the task, sometimes two or three terms are triggered more or less simultaneously by the same prior term, and one must enter them in sequence; in this case the last entered of these terms would be related not to the immediately preceding term but to one a step or two earlier. It is acknowledged that these defining criteria for a jump are somewhat arbitrary, and depend on a domain expert's subjective judgment; given a reference structure for comparison (perhaps formed from several experts' concept maps), a cleaner definition should be possible for follow-up study.

Define a subject's jump rate on an FTE task to be the number of jumps occurring in her response list, divided by the total number of terms in the list. A jump rate of zero would indicate that every term is related to one of the previous three terms; a jump rate of one would indicate that every term was unrelated to all of the previous three.

Each of the sixteen subjects in the study was given a FTE task once, during session J. The specified topic area for terms was "the material covered in Physics 151." (Physics 151 was the current course from which subjects were drawn, and session J was given between the penultimate and ultimate classes of the course.) The task lasted for 30 minutes. The number of responses entered by students during that time ranged from 37 to 174, with a mean of 81 and a standard deviation of 34. Calculated jump rates ranged from 0.18 to 0.60, with a mean of 0.32 and standard deviation of 0.12.

Because inspection of the response lists showed that terms were entered much more sporadically during the later part of the task for all students, with many long pauses, isolated terms, and questionable terms, it was suspected that the earlier portion of the task might better reveal subjects' degree of structuring of the domain concepts, and the latter part might simply introduce noise to the measurement. To investigate this, jump rates were also calculated for the first half of each subject's response list (determined by term count, not by time). First-half jump rates ranged from 0.11 to 0.60, with a mean of 0.26 and a standard deviation of 0.12.

Figure 5 shows a scatterplot of subjects' overall jump rates against the sum of their raw course exam scores, and Fig. 6 shows the same for first-half jump rates. The coefficient of correlation is $r = -0.23$ for Fig. 5 and $r = -0.46$ for Fig. 6, where $\pm 0.13$ is the threshold for statistical significance with 16 data points.

Although the coefficient of correlation for these plots might seem to indicate a significant correlation, the plots show one outlying point (top left) which overly influences the results. Recalculating the coefficient of correlation without this outlier yields $r = 0.16$ for the jump rate and $r = -0.09$ for the first-half jump rate, where $\pm 0.14$ is the threshold for statistical significance with 15 data points. The evidence for the hypothesized correlation is statistically marginal.

In contrast, Fig. 7 shows a plot of FTE jump rate versus course exam performance for a different study, consisting of 18 students from the Fall 1997 Physics 152 course (second-semester introductory physics for science and engineering majors). The designated topic area was "the material covered in Physics 152." One subject was removed from the sample because he or she did not take one of the course exams. Because the exam grades for this course were normalized to a 100-point scale, the measure of overall exam performance used was the average of the four course exam scores. Figure 8 shows the same plot for first-half jump rates. The correlation of coefficient for jump rates versus exam performance is $r = -0.67$, and for first-half jump rates versus exam performance is $r = -0.59$, where $\pm 0.13$ is the threshold for statistical significance with 17 data points.

The data from this study suggest a statistically significant correlation. It is not clear why the results from the two studies seem to differ; perhaps the difference in subject material or student composition of the two courses is relevant, or some aspect of the study itself. Further research is required to resolve the question of whether FTE jump rate correlates with course performance. Two specific questions to investigate are whether an improved definition of "jump" can be found that strengthens the correlation, and whether the jump rate would correlate more strongly with a better measure of course performance than the standard multiple-choice exams used.

## IV. DISCUSSION

Overall, the results presented above indicate that ConMap term entry tasks can elicit information about introductory physics students' conceptual knowledge structure. Comparison of TPTE term lists with drawn concept maps reveals that the TPTE lists approximate a subset of concept map nodes, primarily the most central nodes of the map. Experts' ratings of the "quality" of subjects' TPTE responses correlate with student exam scores. PPTE responses are more likely to include terms indicative of the correct answer to the prompt problem as a result of course instruction on the relevant physics. And FTE jump rates might correlate with exam scores. These results are preliminary and in need of corroboration, but should justify further investigation of the ConMap approach to assessment.

It is worth noting that although correlations with subjects' performance on course exams have been used to validate some of the ConMap measures, the ConMap assessments have in fact been developed in response to the perceived inadequacy of traditional course exams, and are not intended to reproduce exam results. Follow-up research should validate the ConMap tasks against carefully constructed assessments of conceptual expertise, perhaps including performance on hand-crafted "conceptual problems."

The overlap between TPTE responses and HDCM nodes, especially central (level one) nodes, suggests the possibility of using a TPTE-based assessment as an easier-to-administer, easier-to-evaluate equivalent to the much-studied HDCM. In further research, one might compare branchings from subsidiary nodes of a HDCM to response lists when the subsidiary node term is used as a TPTE prompt. It might be possible to predict significant portions of a subject's HDCM from TPTE response data for a set of prompt terms.

The resulting TPTE-elicited network might even be more revealing of true knowledge structure than a consciously drawn map. One might consider whether the fact that some of a subject's level one HDCM terms appear in the TPTE response list and some do not indicates anything fundamental about the subject's knowledge, rather than indicating that the TPTE task is noisy. Perhaps the HDCM level one terms which also appear in the TPTE are those to which the subject has automated, instant access, while those which don't appear are only accessible to the subject upon conscious reflection. Here again, further study is warranted.

The work described in this paper is merely one early step in the achievement of the long-term goals described in the Introduction: the development of practical, efficient assessment tools for diagnosing physics students' evolving conceptual knowledge structures, and the formulation of appropriate detailed cognitive models by which to diagnose. Research towards this goal faces the circularity problem common to all new physics frontiers: an appropriately sensitive experimental probe is a precondition to useful empirical data, yet the data validate a proposed probe; a model is necessary to interpret the data, but the data suggest and constrain modeling; and when a new probe is needed, its invention is guided by some kind of model. Thus, this paper has sought to present some possible probes, along with demonstrations of the kind of empirical data they make possible, and evidence of interpretability in the data which serves to justify the probes. A forthcoming paper[24] will explore the connection to modeling knowledge structure, access, and evolution.

[a]Electronic Mail: beatty@physics.umass.edu

[1]John R. Anderson, *Rules of the Mind* (Lawrence Erlbaum Associates, Hillsdale, NJ, 1993).

[2]Jill H. Larkin, "Information processing models and science instruction," in *Cognitive Process Instruction*, edited by Jack Lochhead and John Clement (The Franklin Institute Press, Philadelphia, Pennsylvania, 1979), pp. 109-118.

[3]Michelene T. H. Chi, Paul J. Feltovich, and Robert Glaser, "Categorization and representation of physics problems by experts and novices," Cog. Sci. **5**, 121-152 (1981).

[4]William J. Gerace, "Contributions from cognitive research to mathematics and science education," presented at the Workshop on Research in Science and Mathematics Education, Cathedral Peak, South Africa, 1992 (unpublished).

[5]Jose P. Mestre, Robert J. Dufresne, William J. Gerace *et al.*, "Promoting skilled problem-solving behavior among beginning physics students," J. Res. Sci. Teach. **30** (3), 303-317 (1993).

[6]William J. Gerace, William J. Leonard, Robert J. Dufresne *et al.*, Report No. PERG-1997#05-AUG#2-26, pp. 1997.

[7]Richard Zajchowski and Jack Martin, "Differences in the problem solving of stronger and weaker novices in physics: Knowledge, strategies, or knowledge structure?," J. Res. Sci. Teach. **30** (5), 459-470 (1993).

[8]Edward F. Redish, "Implications of cognitive studies for teaching physics," Am. J. Phys. **62** (9), 796-803 (1994).

[9]Jose Mestre and Jerold Touger, "Cognitive research — what's in it for physics teachers?," Phys. Teach. **27** (6), 447-456 (1989).

[10]Pamela H. Hardiman, Robert J. Dufresne, and Jose P. Mestre, "The relation between problem categorization and problem solving among experts and novices," Memory and Cognition **17** (5), 627-638 (1989).

[11]J. D. Novak and D. B. Gowin, *Learning How to Learn* (Prentice-Hall, Englewood Cliffs, NJ, 1984).

[12]Maria Araceli Ruiz-Primo and Richard J. Shavelson, "Problems and issues in the use of concept maps in science assessment," J. Res. Sci. Teach. **33** (6), 569-600 (1996).

[13]Michael J. Young, "Quantitative measures for the assessment of declarative knowledge structure characteristics," Ph.D. thesis, University of Pittsburgh, 1993.

[14]Diana C. Rice, Joseph M. Ryan, and Sara M. Samson, "Using concept maps to assess student learning in the science classroom: Must different methods compete?," J. Res. Sci. Teach. **35** (10), 1103-1127 (1998).

[15]Alberto Regis, Pier Giorgio Albertazzi, and Ezio Roletto, "Concept maps in chemistry education," J. Chem. Educ. **73** (11), 1084-1088 (1996).

[16]Tsai-Ping Ju, "The development of a microcomputer-assisted measurement tool to display a person's knowledge structure," Ph.D. thesis, University of Pittsburgh, 1989.

[17]Timothy E. Goldsmith, Peder J. Johnson, and William H. Acton, "Assessing structural knowledge," J. Educ. Psych. **83** (1), 88-96 (1991).

[18]Pilar Gonzalvo, José J. Cañas, and María-Teresa Bajo, "Structural representations in knowledge acquisition," J. Educ. Psych. **86** (4), 601-616 (1994).

[19]N. M. Cooke, F. T. Durso, and R. W. Schvaneveldt, "Recall and measures of memory organization," J. Exper. Psych.: Learning, Memory, and Cognition **12** (4), 538-549 (1986).

[20]C. J. F. ter Braak, "Ordination," in *Data Analysis in Community and Landscape Ecology*, edited by R. H. G. Jongman, C. J. F. Ter Braak, and O. F. R. Van Tongeren (Cambridge University Press, Cambridge, 1995).

[21]Peder J. Johnson, Timothy E. Goldsmith, and Kathleen W. Teague, "Similarity, structure, and knowledge: A representational approach to assessment," in *Cognitively Diagnostic Assessment*, edited by Paul D. Nichols, Susan F. Chipman, and Robert L. Brennan (Lawrence Erlbaum Associates, Hillsdale, NJ, 1995), pp. 221-250.

[22]Ian Beatty, "ConMap: Investigating New Computer-Based Approaches to Assessing Conceptual Knowledge Structure in Physics," Ph.D. thesis, University of Massachusetts, 2000.

[23]N. C. Barford, *Experimental Measurements: Precision, Error and Truth* (John Wiley & Sons, NY, 1967), 2nd ed.

[24]Ian D. Beatty, William J. Gerace, and Robert J. Dufresne, "Probing and modeling physics students' conceptual knowledge structures through term association times," to be submitted.

Table I: Mean and standard deviation across study subjects for the fraction of level 1 HDCM terms appearing in the corresponding TPTE response list, for each of the four HDCM/TPTE sets.

Table II: Mean and standard deviation across study subjects of the fraction TPTE response terms not appearing on the corresponding HDCM, for each of the four HDCM/TPTE sets.

| comparison | mean | standard deviation |
|---|---|---|
| B-HDCM vs. B-TPTE (“force”) | 0.54 | 0.22 |
| G-HDCM vs. J-TPTE (“energy”) | 0.61 | 0.19 |
| H-HDCM vs. H-TPTE (“momentum”) | 0.64 | 0.18 |
| J-HDCM vs. J-TPTE (“force”) | 0.47 | 0.18 |
| All 4 combined | 0.57 | 0.20 |

| comparison | mean | standard deviation |
|---|---|---|
| B-HDCM vs. B-TPTE (“force”) | 0.19 | 0.20 |
| G-HDCM vs. J-TPTE (“energy”) | 0.13 | 0.13 |
| H-HDCM vs. H-TPTE (“momentum”) | 0.13 | 0.14 |
| J-HDCM vs. J-TPTE (“force”) | 0.28 | 0.12 |
| All 4 combined | 0.18 | 0.16 |

Figure 1: Dialog box for *Term Prompted Term Entry* (TPTE) task.

Figure 2: TPTE response list “scores”, averaged over three presentations of the prompt “force”, vs. overall course exam performance.

Figure 3: Prompt problem for PPTE I2 and J4.

Figure 4: Prompt problem for PPTE I5 and J1.

Figure 5: FTE jump rate vs. course exam performance (Physics 151 Spring 1999 study).

Figure 6: FTE jump rate for first half of response list vs. course exam performance (Physics 151 Spring 1999 study).

Figure 7: FTE Jump rate vs. course exam performance (Physics 152 Fall 1997 study).

Figure 8: FTE jump rate for first half of response list vs. course exam performance (Physics 152 Fall 1997 study).

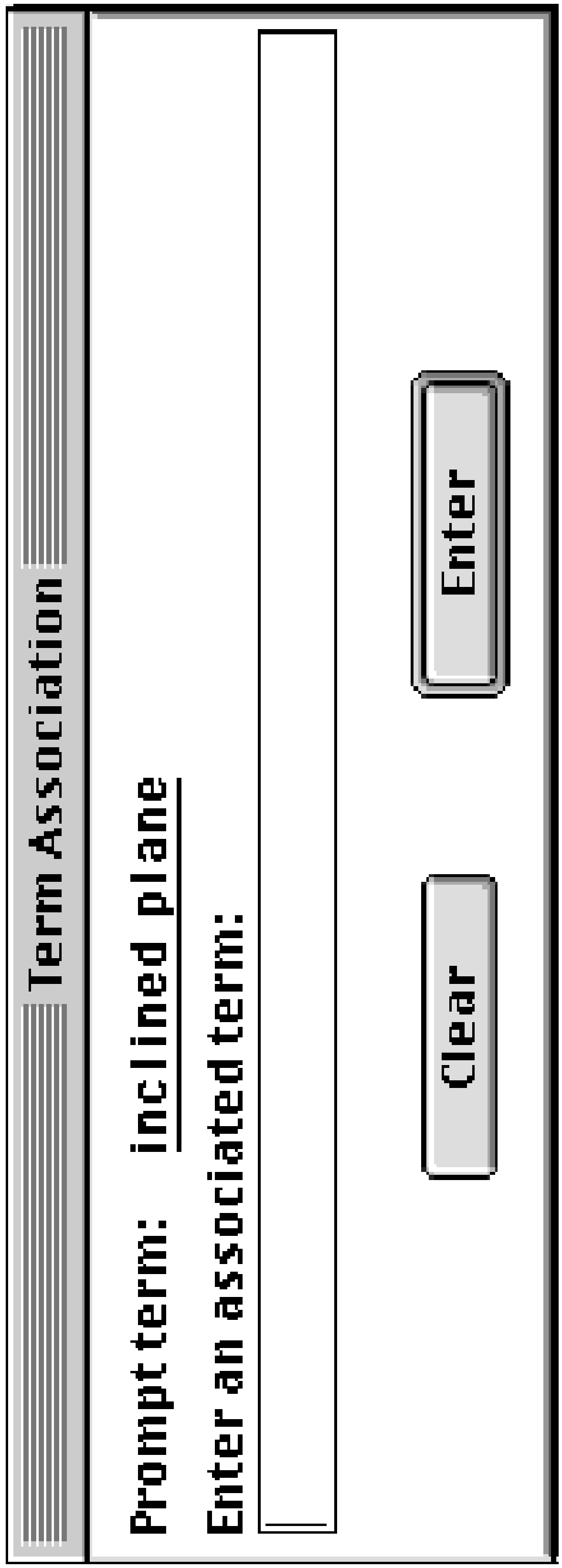
Term Association
Prompt term: inclined plane
Enter an associated term:
Clear
Enter

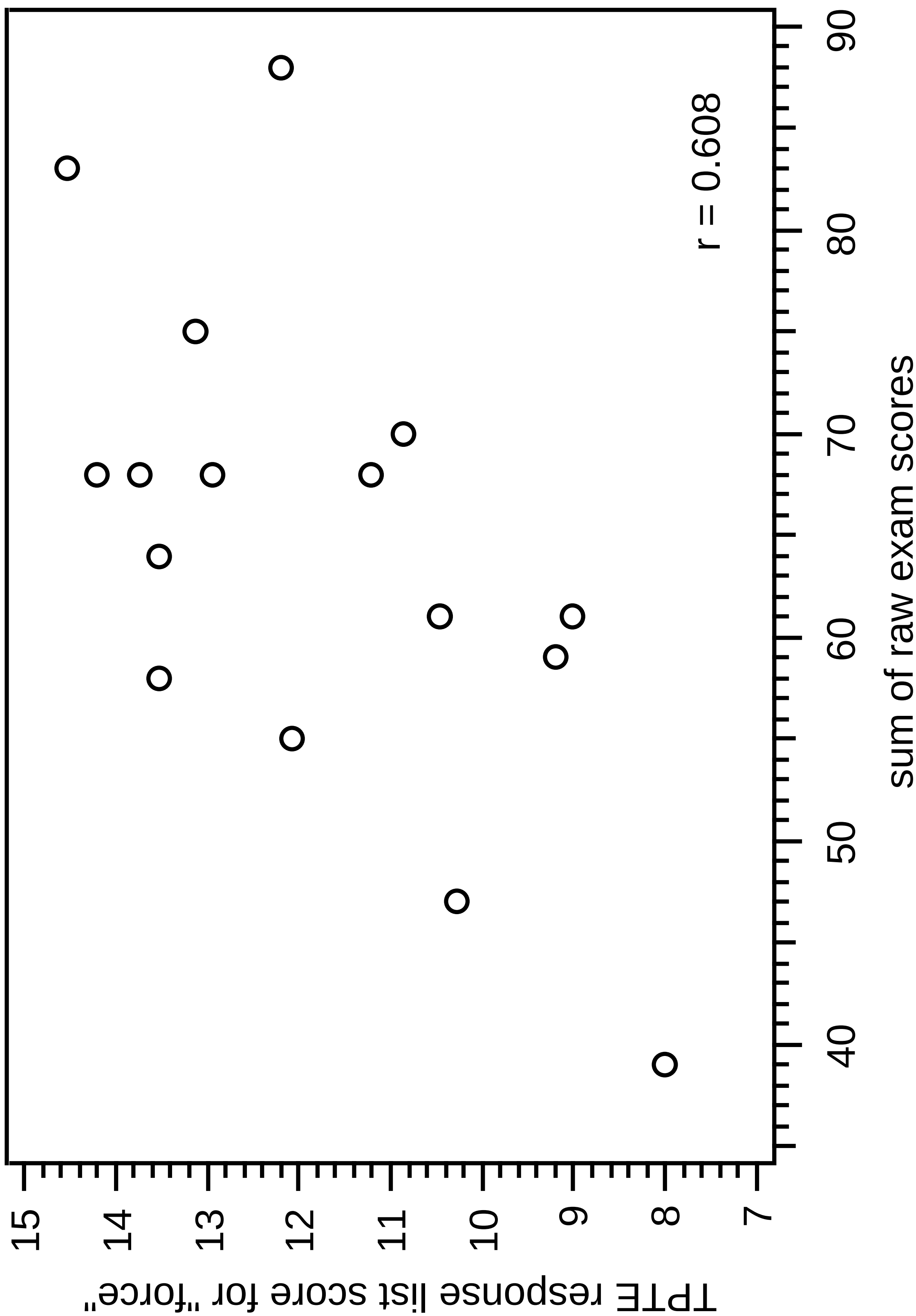
r = 0.608
sum of raw exam scores
40
50
60
70
80
90
TPTE response list score for "force"
7
8
9
10
11
12
13
14
15

A cannon mounted on top of a wagon fires a cannonball horizontally at a muzzle speed of 50 m/s, as shown. The mass of the wagon and cannon is 100 kg, and the mass of the cannonball is 5 kg. The system is initially at rest prior to the cannonball being fired. What is the final speed of the wagon and cannon immediately after the cannonball is fired?

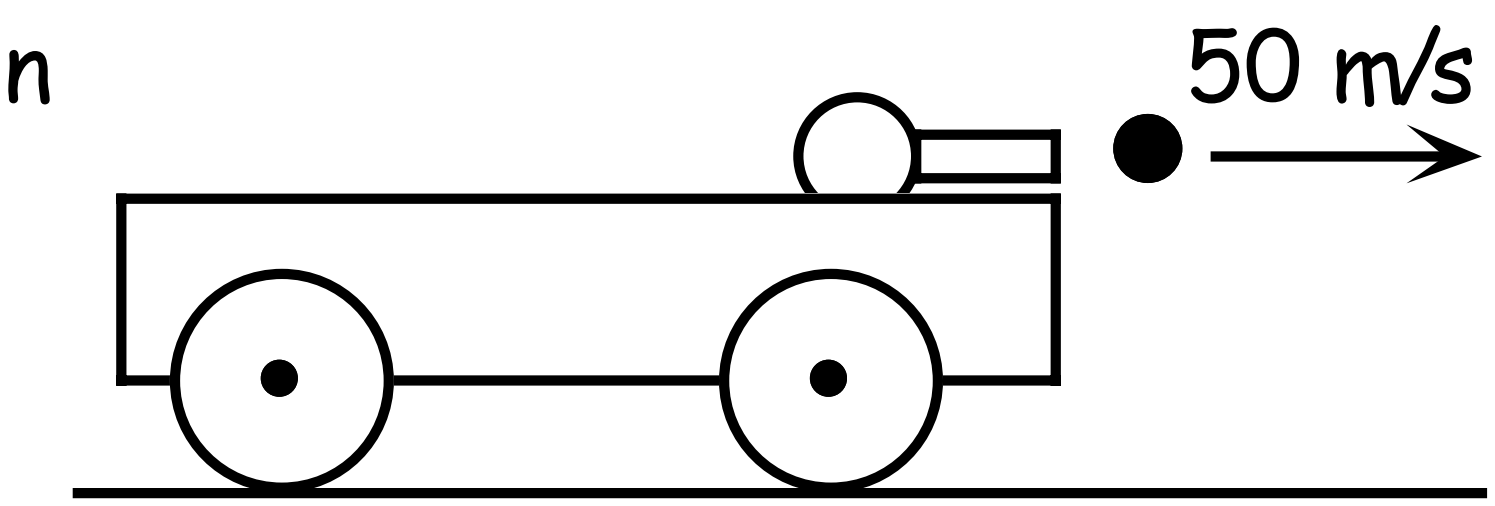

A pendulum is made by attaching a mass of 0.5 kg to a string 1 m long. The pendulum is released from rest with the string horizontal as shown. When the pendulum mass gets to the bottom of the swing, it collides, and sticks to, another mass of 1.5 kg. How high above the ground do the two masses rise after the collision?

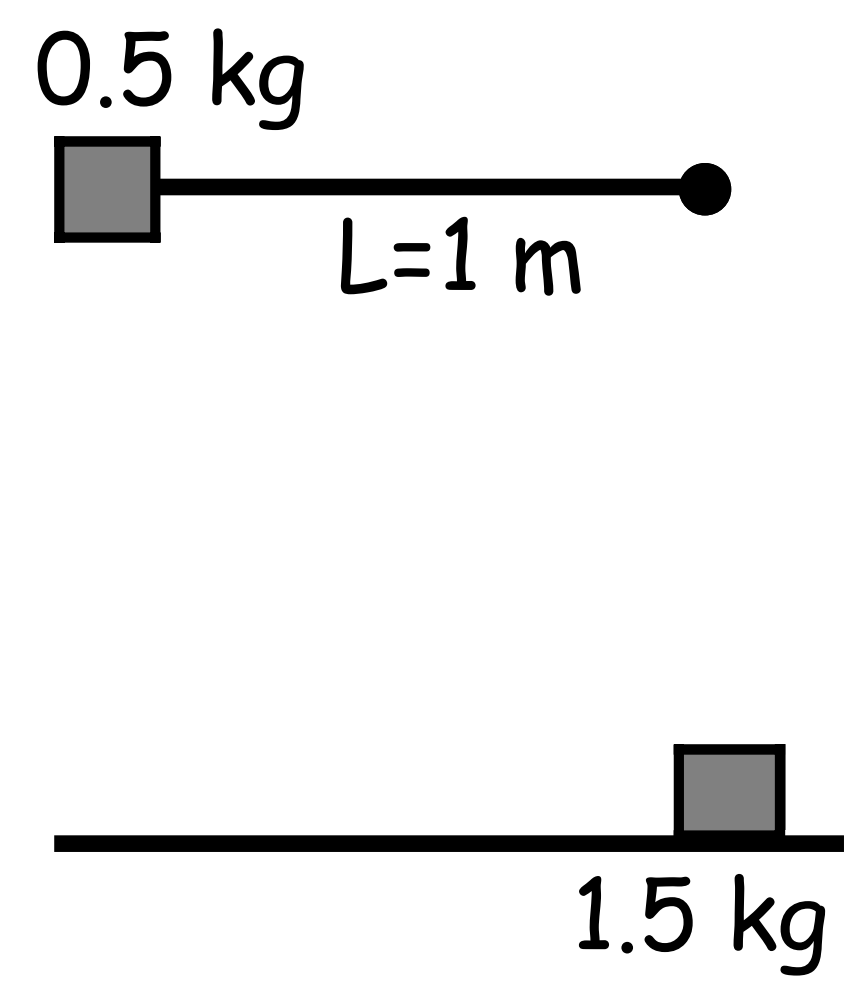

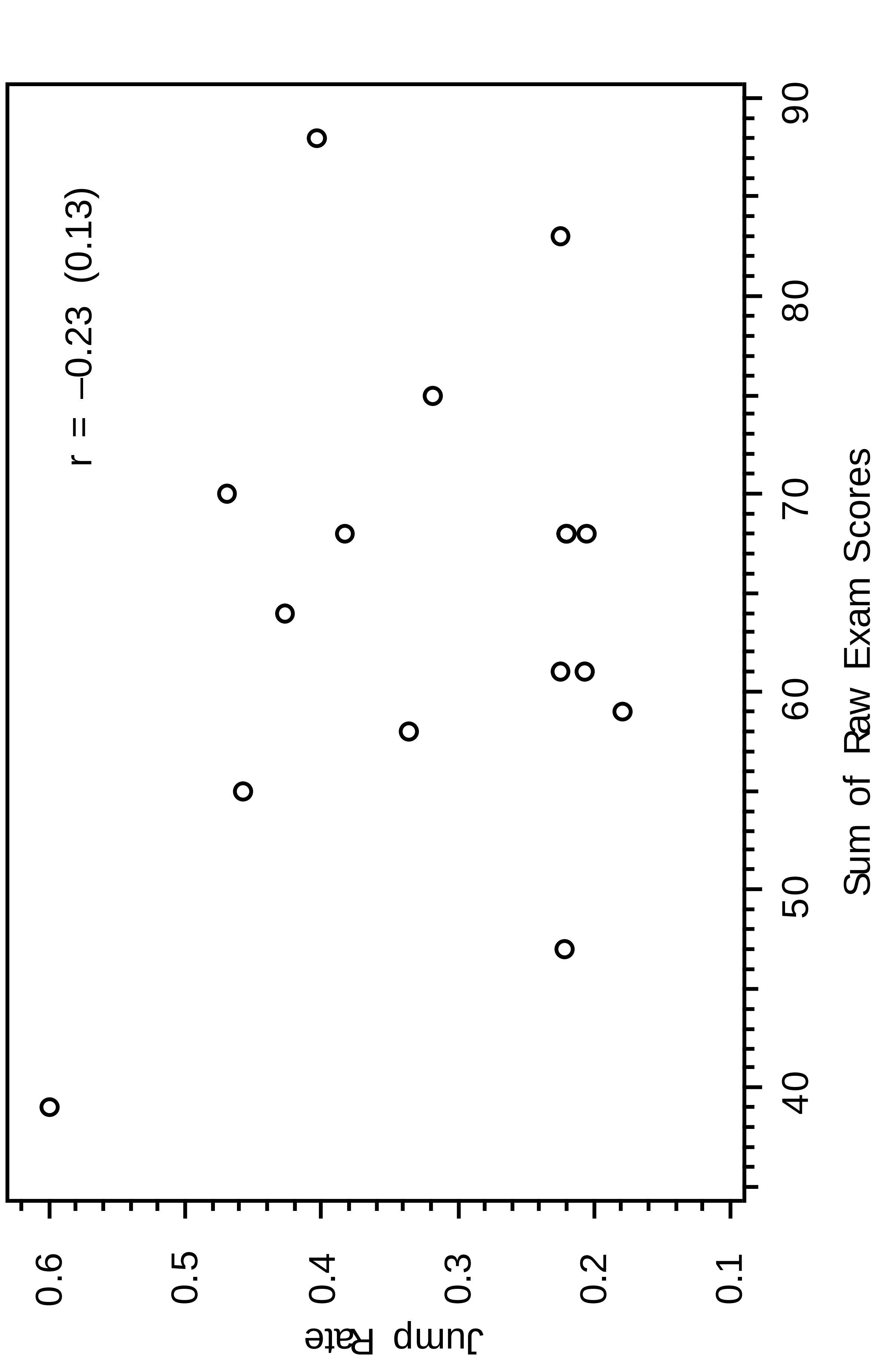
r = –0.23 (0.13)
Jump Rate
0.6
0.5
0.4
0.3
0.2
0.1
Sum of Raw Exam Scores
40
50
60
70
80
90

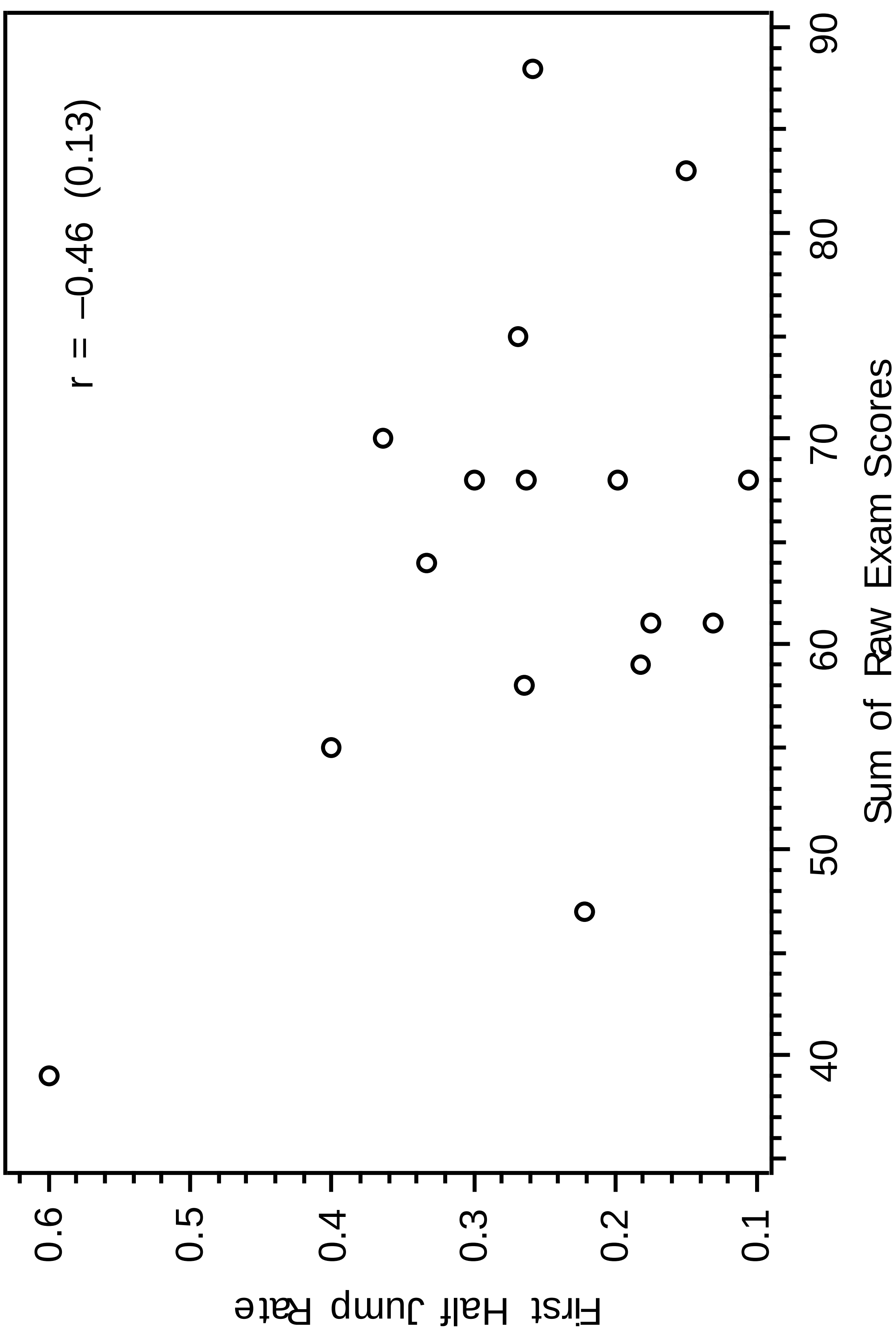
r = –0.46 (0.13)
Sum of Raw Exam Scores
40
50
60
70
80
90
First Half Jump Rate
0.6
0.5
0.4
0.3
0.2
0.1

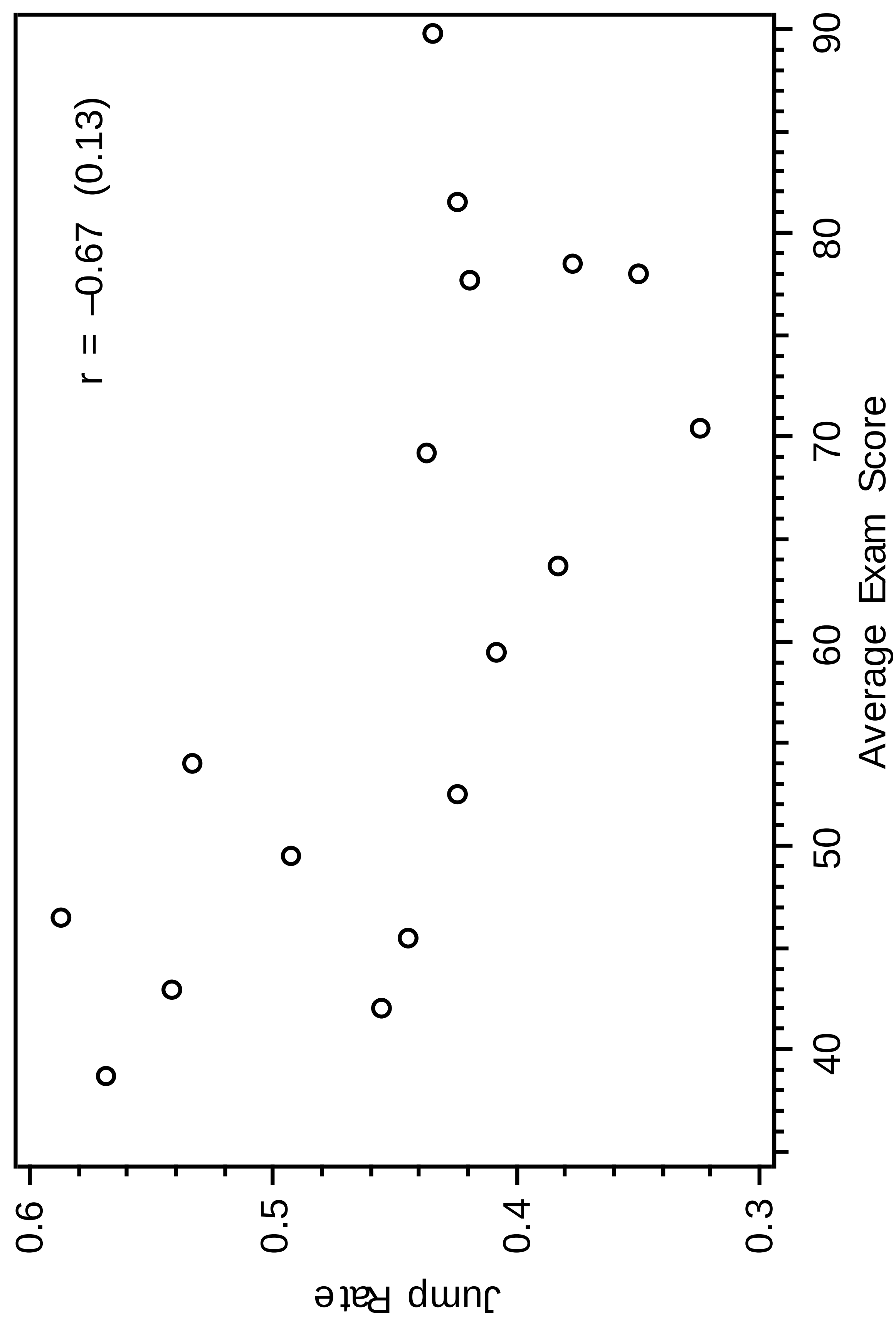
r = –0.67 (0.13)
Average Exam Score
Jump Rate
40
50
60
70
80
90
0.6
0.5
0.4
0.3

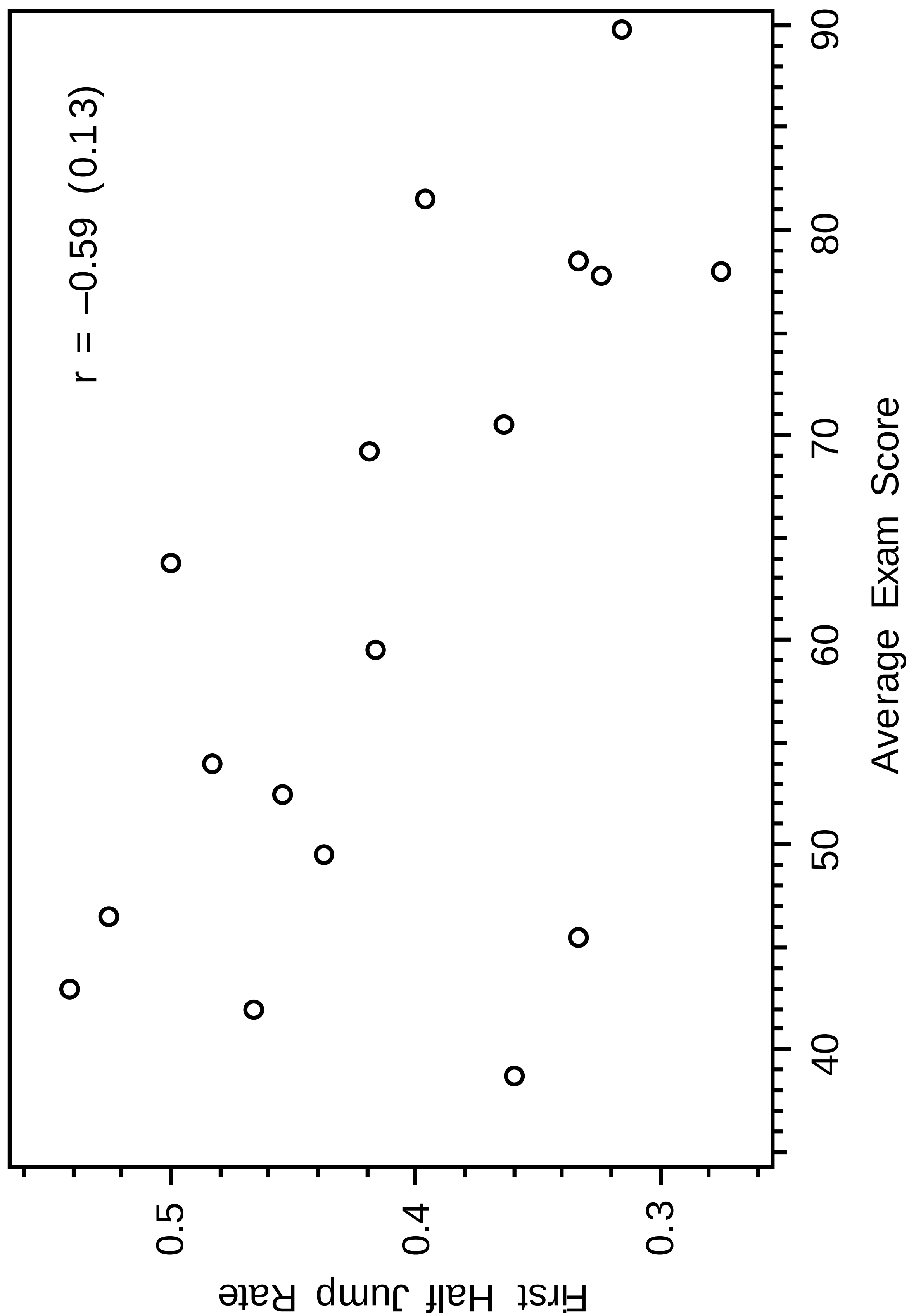
r = –0.59 (0.13)
Average Exam Score
40
50
60
70
80
90
First Half Jump Rate
0.5
0.4
0.3